\documentclass[english,11pt]{article}
\usepackage[latin9]{inputenc}
\usepackage{xcolor}
\usepackage{pdfcolmk}
\usepackage{float}
\usepackage{amsthm}
\usepackage{amsmath}
\usepackage{graphicx}
\usepackage{setspace}
\usepackage{esint}
\PassOptionsToPackage{normalem}{ulem}
\usepackage{ulem}
\makeatletter

\floatstyle{ruled}
\newfloat{algorithm}{tbp}{loa}
\providecommand{\algorithmname}{Algorithm}
\floatname{algorithm}{\protect\algorithmname}
\providecolor{lyxadded}{rgb}{0,0,1}
\providecolor{lyxdeleted}{rgb}{1,0,0}
\numberwithin{equation}{section}
\numberwithin{figure}{section}
\floatstyle{ruled}
\newfloat{algorithm}{tbp}{loa}
\floatname{algorithm}{Algorithm}
\usepackage[noend]{algorithmic}

\newcommand{\ifbody}[1]{ #1  \ENDIF}
\newcommand{\whilebody}[1]{ #1  \ENDWHILE}

\@ifundefined{date}{}{\date{}}
\usepackage{geometry}

\@ifundefined{showcaptionsetup}{}{%
 \PassOptionsToPackage{caption=false}{subfig}}
\usepackage{subfig}
\makeatother

\usepackage{babel}
\begin{document}

\title{A Bayesian approach to accounting for variability in mechanical properties in biomaterials}

\author{Srikrishna Doraiswamy, Arun R.~Srinivasa }

\maketitle
\begin{abstract}
In this paper, we present an approach for modeling biotissues that incorporates the variability in properties as part of their characteristics. This is achieved by considering the parameters of the model of a biomaterial to themselves be random variables and represented by a probability distribution over the space of parameters. This probability distribution is obtained by the systematic use of Bayesian inference together with a continuum mechanics based solution of a boundary value problem. We illustrate this approach by characterizing sheep arteries by using a combination of experimental data and different hyperelastic models. Furthermore, we also develop a model based Bayesian classification of new data into different classes based on the computed model parameter probability distribution.
\end{abstract}
\section{Introduction\label{sec:Introduction}}

Several continuum mechanics models have been proposed to characterize and predict the mechanical response of biomaterials such as blood vessels, the heart and bones \cite{mollica_modeling_2007,humphrey_cardiovascular_2002,fung_biomechanics:_1993,holzapfel_mechanics_2006}. The major challenge in the use of such models is to identify the model parameters to match experimental data. Conventionally, the least squares approach has been used to identify such model parameters. Unlike metallic/manufactured materials, it is not possible to enforce control in composition, the size and shape of test specimens when considering biomaterials. Neither is it possible to test a large number of samples of closely matching characteristics.  Indeed, the variance observed in experiments across different samples reflects this lack of control (see, for example, the data reported in \cite{van_andel_mechanical_2003,vande_geest_effects_2006,garcia-herrera_mechanical_2012,carboni_passive_2007,holzapfel_determination_2005}). Furthermore, the quality of available experimental data also limits the ability of models in accurately representing and predicting the behavior of biomaterials, i.e. there is substantial epistemic uncertainty in the models. Nevertheless such models are useful in characterizing the behavior of such materials. Given the uncertainties both due to lack of a perfect model as well as those due to the experiments and specimens, we propose to replace a deterministic idea of model parameters with a probability distribution for the model parameters. Before we proceed, we observe that a given continuum model with unknown parameters represents a parametrized class of models. Our approach is to characterize groups of samples as probability distributions on the space of model parameters. This poses two key questions that we seek to answer, especially with regards to nonlinear models. For example, 
\begin{enumerate}
\item Given sample data and a putative model with certain model parameters, how does one find a probability distribution for model parameters? If one treats model parameters as being deterministic values, this question can be answered very easily by means of a least squares fit. However, such an approach may not represent any of the samples, nor does it capture the variability that is inherent in the samples.
\item When new samples are provided, how does one determine whether this sample belongs to the same model class as before or is it different? This is a fundamental question that needs to be answered in all diagnostic activities.
\end{enumerate}
We answer these questions by a systematic use of a Bayesian inference approach to both questions (1) and (2). We will illustrate the approach by considering data for arteries reported in \cite{paul_andersohn_modeling_2013}. The Bayesian approach utilized here also circumvents issues of non-uniqueness in least squares estimates (see \cite{ogden_fitting_2004} for a discussion on non-uniqueness in least squares estimates for hyperelastic models). The method presented here comes with a cost however. Calculating probability distributions on parameter spaces requires the repeated solution of boundary value problems for different values of parameters as part of Markov Chain Monte Carlo scheme for sampling a probability distribution. Hence, the availability of exact solutions is of vital importance in carrying out efficient simulations.

\subsection*{Key results from this approach}
\begin{enumerate}
\item We provide a methodology that allows us to find probability distributions for model parameters from experimental data. The results of the application of this approach to the characterization of sheep arteries based on cylindrical inflation data, shows the efficacy of the approach in identifying likely ranges of parameters.
\item We present a novel way of calculating the pressure vs. radius response, without the use of specific constitutive equations and the momentum balance equations. Instead, the pressure vs. radius response is obtained by a minimization process that allows us to postpone the choice of a specific constitutive equation. We are thus able to quickly and efficiently obtain the model predictions for both isotropic and orthotropic materials without having to re-solve a boundary value problem.
\item We also provide a methodology for probabilistic classification of samples that can be used to answer the question: What is the probability that a particular sample belongs to a particular model class? The results also highlight the importance of proper priors on the model parameters.
\end{enumerate}

\subsection{Organization of the paper}

The paper is organized into the following 5 sections. Section \eqref{sec:Experimental-Data} deals with the data from cylindrical inflation of sheep aorta samples. Section \ref{sec:Model-for-artery} details the solution of the artery inflation problem based on the assumption that the response can be characterized by using an homogeneous, incompressible, hyperelastic, isotropic solid%
\footnote{Clearly, these assumptions are far too idealistic. However, given the limited data, we want to highlight the core idea of parameter estimation and so we have deliberately not used any other additional information that could have been obtained by knowledge of the structure of the artery. We will demonstrate that these assumptions are sufficient for purposes of classification.%
}. An analytical solution is obtained for different nonlinear models, all of which are represented by two parameters. Thus we seek probability distributions over a two dimensional parameter space for each of these models. Section \ref{sec:Probabilistic-Framework} introduces the Bayesian probabilistic framework used for obtaining the model parameter probability distributions. Section \ref{sec:Results-and-Discussion} lists the details of the algorithm and simulations performed in order to obtain the probability distribution. Section \ref{sec:Using-model-parameter} demonstrates how the probability that is obtained from the ``training'' data can be used to classify ``new'' measurements.

\section{Experimental Data\label{sec:Experimental-Data}}

The experiments used for demonstrating the classification algorithm presented in this work were recently conducted \cite{paul_andersohn_modeling_2013} at Texas A\&M University. Three inflation experiments were performed on 5 samples of sheep aorta at the axial stretch corresponding to \emph{in vivo }axial stretch. The details of the experimental protocol can be found in\cite{paul_andersohn_modeling_2013}. The data was reported as applied pressure (in mmHg) vs.~volume (in mL). This data was reduced to pressure vs.~internal radius (in mm). The data, in the reduced form is shown in figure \ref{fig:Inflation-experimental-data}. The data reveals considerable amount of spread even in such controlled samples. Furthermore, a close look at the figure reveals that there appears to be two different responses among the samples. We will later use this preliminary observation as the basis for the classification problem in section\ref{sec:Using-model-parameter} .

\begin{center}
\begin{figure}[H]
\begin{centering}
\includegraphics[scale=0.75]{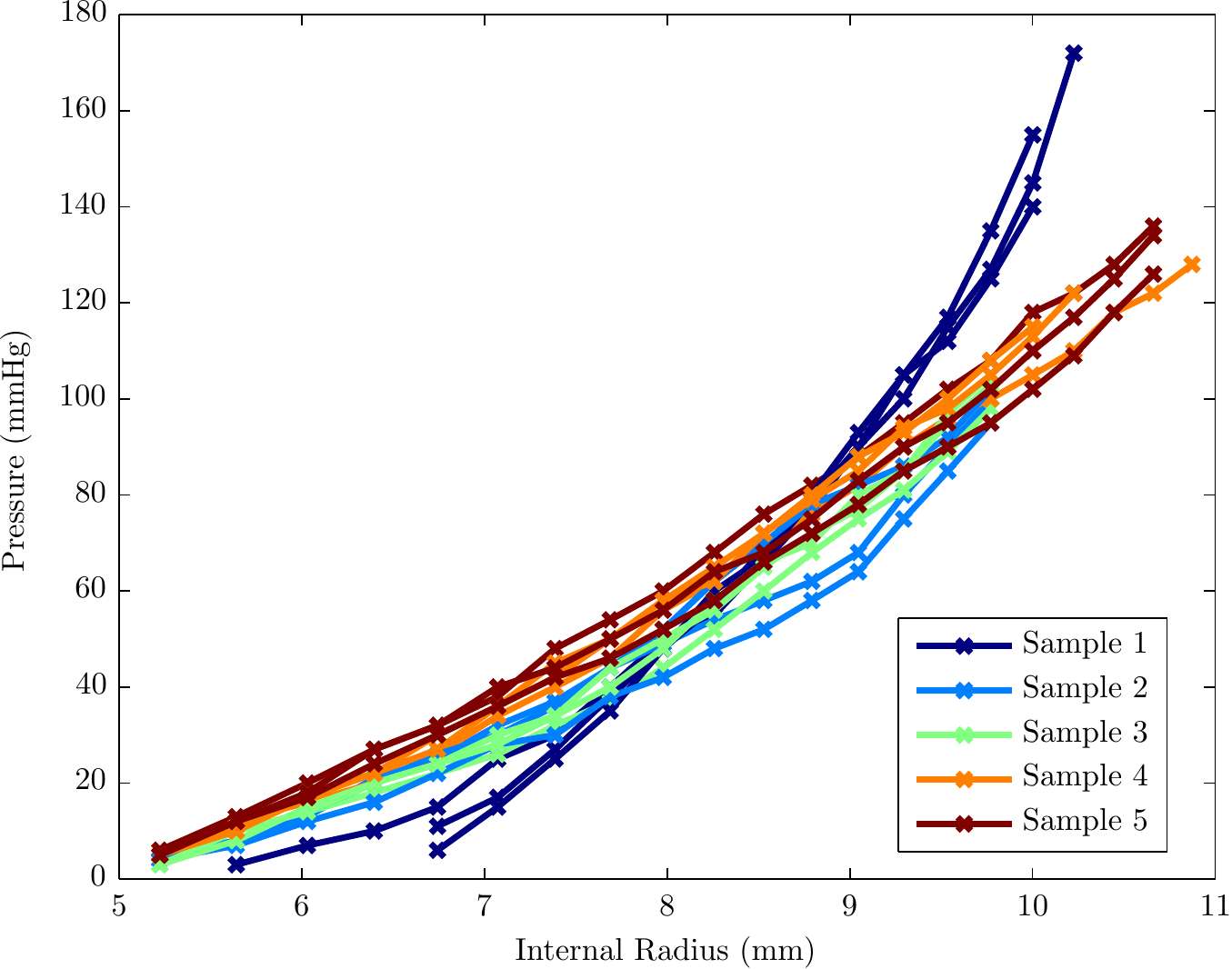}
\par\end{centering}

\caption{Inflation experimental data from 5 samples of sheep aorta -- Pressure vs. Internal radius \cite{paul_andersohn_modeling_2013}. Note that in addition to a significant variance in the reported measurements of different experiments for a given sample, there is also a variance in the behavior across samples.\label{fig:Inflation-experimental-data}}
\end{figure}

\par\end{center}

\section{Model for artery inflation\label{sec:Model-for-artery}}

 For the purposes of this work the artery is modeled as a thick-walled tube made of a homogeneous, incompressible, and hyperelastic material. This choice is motivated by the fact that the exact geometry of the arteries that were tested is not reported, and furthermore the data is presented as (section \eqref{sec:Experimental-Data}) the applied pressure vs. internal volume (and reduced to pressure vs. internal radius). The model that we develop is motivated towards obtaining the model predictions of the applied pressure $p_{\mathrm{mod}}(\boldsymbol{\theta},r_{i})$ as a function of the internal radius $r_{i}$ and the model parameters $\boldsymbol{\theta}$ of the strain energy chosen. The problem statement for the inflation problem and the solution technique used for the solving this problem are explained below.

\subsection{Problem statement}

Consider a thick walled cylinder $B$ that occupies the reference configuration $k_{r}(B)$ which in polar coordinates is represented by the region

\begin{equation}
\Omega_{0}=\{(R,\Theta,Z\}\mid A\leq R\leq B,0\leq\Theta\leq2\pi,0\leq Z\leq L\}.
\end{equation}

Let $\mathbf{X}\in k_{r}(B)$ denote the position vector of a typical particle in the body. At time $t$, the body occupies a configuration $k_{t}(B)$ such that position of the same particle is now $\mathbf{x}$. The motion of the body can be described using the function $\boldsymbol{\chi}_{k_{r}}$

\begin{equation}
\mathbf{x}=\boldsymbol{\chi}_{k_{r}}(\mathbf{X},t)
\end{equation}

The deformation gradient $\mathbf{F}$ and the right Cauchy-Green stretch tensor are given respectively by

\begin{equation}
\mathbf{F}=\frac{\partial\boldsymbol{\chi}_{k_{r}}}{\partial\mathbf{X}},\mathbf{C}=\mathbf{F}^{\mathrm{T}}\mathbf{F}
\end{equation}

Consider the inflation of this annular region under an applied pressure on the internal surface $R=A$%
\footnote{Given the fact that it is the internal radius that can be inferred from the experimental data, we will write all our equations in terms of this parameter $A$%
}. Under this applied pressure, the deformation of the cylinder is given by 

\begin{align}
r & =f(R),\theta=\Theta,z=\lambda_{z}Z\label{eq:Deformation}
\end{align}

where it is assumed that the current radius of the cylinder is independent of $\theta$ and $z$.

\subsection{Constitutive relation}

For the deformation as stated in \eqref{eq:Deformation} the components of the deformation gradient in polar coordinates is given by

\begin{align}
F & =\mathrm{diag}\left(\frac{\mathrm{d}f(R)}{\mathrm{d}R},\frac{f(R)}{R},\lambda_{z}\right)
\end{align}

Enforcing the incompressibilty constraint in the form $\det\mathbf{F}=1$, we obtain the well-known result,

\begin{equation}
r=f(R)=\sqrt{\frac{R^{2}+2c\lambda_{z}}{\lambda_{z}}},\label{eq:FuncOfR}
\end{equation}

The function $f(R)$ contains one constant of integration, $c$. Traditionally this constant is evaluated by using information from the geometry in reference and current configurations (e.g.\ $R=A,\,\lambda_{z}=1,\, r=a,\,\Rightarrow c=\frac{a^{2}-A^{2}}{2}$). We will depart from this approach, and retain this constant of integration $c$. Note that the constant $c$ controls the inflated radius of the cylinder and is consequently a function of the applied pressure.

For future reference, the principal stretches for this deformation are given by

\begin{subequations}\label{eq:Stretches}

\begin{align}
\lambda_{1} & =\frac{\mathrm{d}f(R)}{\mathrm{d}R}=\frac{R}{\lambda_{z}\sqrt{2c+\frac{R^{2}}{\lambda_{z}}}}\\
\lambda_{2} & =\frac{f(R)}{R}=\frac{\sqrt{2c+\frac{R^{2}}{\lambda_{z}}}}{R}\\
\lambda_{3} & =\lambda_{z}
\end{align}

\end{subequations}

and the invariants $\mathrm{I^{C},II^{C}}$ of the right Cauchy-Green stretch tensor are given as

\begin{subequations}\label{eq:Invariants}

\begin{align}
\mathrm{I^{C}} & =\lambda_{1}^{2}+\lambda_{2}^{2}+\lambda_{3}^{2} & =\lambda_{z}^{2}+\frac{R^{2}}{\lambda_{z}\left(R^{2}+2c\lambda_{z}\right)}+\frac{R^{2}+2c\lambda_{z}}{R^{2}\lambda_{z}}\\
\mathrm{II^{C}} & =\lambda_{1}^{2}\lambda_{2}^{2}+\lambda_{2}^{2}\lambda_{3}^{2}+\lambda_{3}^{2}\lambda_{1}^{2} & =\frac{1}{\lambda_{z}^{2}}+\frac{R^{2}\lambda_{z}}{R^{2}+2c\lambda_{z}}+\frac{\lambda_{z}\left(R^{2}+2c\lambda_{z}\right)}{R^{2}}
\end{align}

\end{subequations}

In order to obtain a relationship between the applied pressure and the constant $c$ we employ a strategy, that allows us to obtain a general expression without the necessity for specifying the constitutive equation yet. We do this by noting that the solution to the elastostatics problem, is an extremum of the total potential energy of the system. In other words, the principle of minimum potential energy states that admissible deformations are those that render the potential energy functional to be a extremum. 

For the problem under consideration, the potential energy $\psi$ is, 

$ $
\begin{equation}
\psi=\int_{\Omega_{0}}W\,\mathrm{d}V-p\Delta V\label{eq:PotentialEnergy}
\end{equation}

where $W$ is the strain energy , P is the pressure applied to the surface $R=A$ of the cylinder and $\Delta V$ is the change in volume enclosed by this surface. 

Rewriting equation \eqref{eq:PotentialEnergy}, 

\begin{equation}
\psi=\int\limits _{0}^{L}\int\limits _{0}^{2\pi}\int\limits _{A}^{B}W\, R\,\mathrm{d}R\,\mathrm{d}\Theta\,\mathrm{d}Z-p\times\pi(f(A)^{2}-A^{2})L\,.\label{eq:PotentialEnergy_Simplify_1}
\end{equation}

Since the strain energy is a function of the deformation gradient, which in this case is only a function of $c$, the integrand is a function of $R$ and $c$ only.  Given that the experiments are all conducted at a constant axial stretch ($\lambda_{z}=1$), the only variable in the above equation is $c$. Therefore, since the minimum of the potential energy is attained when the gradient is zero, 

\begin{align}
\frac{\mathrm{d}\psi}{\mathrm{d}c} & =0\Rightarrow p=\int\limits _{A}^{B}\frac{\mathrm{d}W}{\mathrm{d}c}\, R\,\mathrm{d}R\label{eq:PressureVsC}
\end{align}

Equation \eqref{eq:PressureVsC} is the equation that relates the pressure $p$ to the constant of integration $c$ irrespective of the form of the strain energy function. The relation between the pressure and the (current) internal radius is sought since the data is also in this format. Since the relation between $c$ and the internal radius is given by (with $\lambda_{z}=1$)

\begin{equation}
c=\frac{r_{i}^{2}-A^{2}}{2}\label{eq:ExprC}
\end{equation}

The above relation will allow us to calculate the relationship between the pressure and the internal radius as and when required. The specific form of equation \eqref{eq:PressureVsC} depends on the choice of the strain energy function. In contrast to the approach presented in Ogden \cite{ogden_non_1997} which is specific to isotropic materials, the result \eqref{eq:PressureVsC} is valid for any form of the strain energy function, so long as the deformation gradient is acceptable. Thus we can use the expression \eqref{eq:PressureVsC} for the case of an orthotropic material, as would be seen subsequently. 

In the following section, three different strain energies will be used to get the corresponding pressure vs. $c$ relations. Of these, the first two namely the power-law neo-Hookean model and the Ogden model are isotropic whereas the Criscione-type model is that for an orthotropic material.

\subsection{Forms of constitutive relations for specific strain energies\label{sub:Forms-of-constitutive}}

\subsubsection{\uline{Power-law neo-Hookean model}}

The first model we choose for comparison with experimental data is the power-law neo-Hookean, which has been used in the literature for modeling biotissues (e.g. \cite{raghavan_toward_2000}). The strain energy density for the incompressible power-law neo-Hookean model is given by 

\begin{equation}
W=\frac{\mu}{2}\left\{ \left[1+\frac{1}{n}(\mathrm{I^{C}}-3)\right]^{n}-1\right\} ,\label{eq:GeneralizedNeoHookeanModel}
\end{equation}
where $\mu>0$ is the shear modulus and $n$ is a positive number. Note that when $n=1$, the strain energy reduces to that of neo-Hookean model. Substituting for W and $\mathbf{I^{\mathrm{C}}}$ in \eqref{eq:PressureVsC} , we obtain the following equation for constant the pressure in terms of constant $c$, which in turn can be written in terms of the internal radius $r_{i}$ \eqref{eq:ExprC}.

\begin{equation}
p=\int\limits _{A}^{B}\frac{4\mu cn\left(c+R^{2}\right)\left(1+\frac{4c^{2}}{2cnR^{2}+nR^{4}}\right)^{n}}{\left(2c+R^{2}\right)\left(4c^{2}+nR^{2}\left(2c+R^{2}\right)\right)}R\,\mathrm{d}R\label{eq:ConstEqnGenNH}
\end{equation}
The model parameters for this strain energy are $\mu$ and $n$ and we need to obtain a probability distribution on these two parameters.

\subsubsection{\uline{Ogden model}}

The most widely implemented model in the computational literature for incompressible isotropic materials is the Ogden model. In our case, we will consider a two parameter Ogden model given by the strain energy density,

\begin{equation}
W=\frac{2\mu}{\alpha^{2}}\left(\lambda_{1}^{\alpha}+\lambda_{2}^{\alpha}+\lambda_{3}^{\alpha}+\frac{1}{\lambda_{1}^{\alpha}}+\frac{1}{\lambda_{2}^{\alpha}}+\frac{1}{\lambda_{3}^{\alpha}}-6\right)\label{eq:OgdenModel}
\end{equation}

where $\mu$ and $\alpha$ are material constants. The pressure vs. $c$ relation using this strain energy is , 

\begin{align}
C_{i}p & =\int\limits _{A}^{B}-\frac{2\mu\left(\left[\left(\frac{R}{\sqrt{2c+R^{2}}}\right)^{\alpha}-\left(\frac{\sqrt{2c+R^{2}}}{R}\right)^{\alpha}\right]+\left[\left(\frac{\sqrt{2c+R^{2}}}{R}\right)^{-\alpha}-\left(\frac{R}{\sqrt{2c+R^{2}}}\right)^{-\alpha}\right]\right)}{\left(2c+R^{2}\right)\alpha}R\,\mathrm{d}R\label{eq:ConstEqnOgden}
\end{align}

\subsubsection{\uline{Criscione-type model with logarithmic strain attributes}}

The above constitutive equations are based on assuming that the material is isotropic. On the other hand, there is compelling evidence to suggest that artery is orthotropic. However, conventional models of incompressible, orthotropic, hyperelastic materials, cannot be used to find parameters because they are a function of 6 variables that, for the cylinder inflation problem, vary with deformation. Thus, it is impossible to obtain any simple expression for the strain energy. 

Recently however, Criscione \cite{criscione_constitutive_2004} has a developed a modeling approach for orthotropic hyperelastic materials that considerably simplifies this problem by utilizing a set of non-polynomial invariants whose derivatives are almost transversal (see also \cite{srinivasa_use_2012} in this regard). For the special case of cylindrical inflation Criscione has shown that it is possible to have an expression for the strain energy in which only one of the invariants varies with inflation thus allowing for the possibility of obtaining probability distributions for the parameters without full knowledge of the function.

The Criscione form of the strain energy function $W(\gamma_{1},\gamma_{2},\gamma_{3},\gamma_{4},\gamma_{5},\gamma_{6})$ is a function of 6 strain invariants given by,

\begin{align*}
\gamma_{1}=\log J &  & \gamma_{2}=(3/2)\log\lambda_{3} &  & \gamma_{3}=2\log\lambda_{2}+\log\lambda_{3}\\
\gamma_{4}=\phi_{RZ} &  & \gamma_{5}=\phi_{\Theta Z} &  & \gamma_{4}=\phi_{\Theta R}
\end{align*}

where, $J=\det\, F$, and $\phi_{RZ}$, $\phi_{\Theta Z}$, $\phi_{\Theta R}$ are the shear strains in the planes denoted by the corresponding subscripts.

Given the nature of the deformation, for the problem at hand, $\gamma_{3}$ is the only nonzero attribute and so, the chosen form of $W$ is

\begin{equation}
W(\gamma_{1},\gamma_{2},\gamma_{3},\gamma_{4},\gamma_{5},\gamma_{6})=\hat{W}(\gamma_{3})=a\cosh(b\gamma_{3})\label{eq:CriscioneModel}
\end{equation}

The pressure, from \eqref{eq:PressureVsC} is therefore,

\begin{equation}
p=\int\limits _{A}^{B}\enskip\frac{2ab\sinh\left[2b\log\left[\frac{\sqrt{2c+R^{2}}}{R}\right]\right]}{2c+R^{2}}R\,\mathrm{d}R\label{eq:ConstEqnCriscione}
\end{equation}

The model parameters in this model are $a$ and $b$.

With the pressure $p$ expressed as a function of the model parameters, the conventional approach to `fit' the model to data would be to solve a non-linear least squares problem. In general, such an approach is not guaranteed to give a unique solution, because the cost function is not convex. Furthermore, no information about robustness of the solution can be obtained easily. We will depart from this approach and seek probability distributions for the reasons explained in section \eqref{sec:Introduction}.

\section{Probabilistic Framework -- Bayesian Inference \label{sec:Probabilistic-Framework}}

Bayesian inference is the method of using the Bayes' rule to update the state of information using observations. The state of information before the observations were made is known as the \emph{prior }and the updated state of information is known as the \emph{posterior.} The following notation is used in the remainder of the section: let $\mathbf{p}_{\mathrm{exp}}$ be the vector of measured pressure values corresponding to the vector of internal radius values $\mathbf{r}_{i}$. Let $\mathbf{p}_{\mathrm{mod}}(\boldsymbol{\theta})$ be the vector of model predictions corresponding to $\mathbf{r}_{i}$ as a function of the model parameters $\boldsymbol{\theta}$. We are interested in computing the probability distribution $P(\boldsymbol{\theta}\mid\mathbf{p}_{\mathrm{exp}}$), i.e. the probability distribution on the model parameters $\boldsymbol{\theta}$ given the data $\mathbf{p}_{\mathrm{exp}}$.

\subsection{Probability distribution of model parameters\label{sub:Probability-distribution-of}}

As mentioned in the previous section, the classification problem requires the computation of the probability distribution of the model parameters for each of the classes. As mentioned in section \ref{sec:Experimental-Data} the data is reported as pressure, $\mathbf{p}_{\mathrm{exp}}$ vs internal radius $\mathbf{r}_{i}$. The probability distribution of the parameters $\boldsymbol{\theta}$ , given the observations $\mathbf{p}_{\mathrm{exp}}$ is computed using Bayes' rule:

\begin{equation}
P(\boldsymbol{\theta}\mid\mathbf{p}_{\mathrm{exp}})=\frac{P(\mathbf{p}_{\mathrm{exp}}\mid\boldsymbol{\theta})P(\boldsymbol{\theta})}{P(\mathbf{p}_{\mathrm{exp}})}\label{eq:BayesRule_modelParams}
\end{equation}

The prior probability distribution for the parameters $P(\boldsymbol{\theta})$ may be chosen according to any available prior information on the value of the parameters. If there is no distinguishing prior information, then a \emph{non-informative }prior may be chosen so as to not bias the inference process. For the work presented in this paper, a non-informative prior is assumed, resulting in this case a uniform distribution for $P(\boldsymbol{\theta})$.

A standard way of choosing a likelihood function is to assume that it is a Gaussian distribution on the errors which are assumed to be uncorrelated. However, a careful look at the experimental data shows, that the deviations from the model are correlated deviations. Furthermore, we expect the model to capture not only the trend in the pressure, but also the trend in the derivative of the pressure with respect to the radius, since the latter is a measure of local wave speeds in the tissue. Since wave speeds are critical \cite{gosling_terminology_2003} to inference problems based on elastography, it is important to represent the wave speeds accurately also in the model. To this end, let $\mathbf{p}_{\mathrm{mod}}(\boldsymbol{\theta})$ be the model prediction for parameter vector \textbf{$\boldsymbol{\theta}$}, the absolute error between the data and the model is given by $\left\Vert \mathbf{p}_{\mathrm{mod}}(\boldsymbol{\theta})-\mathbf{p}_{\mathrm{exp}}\right\Vert $. The error between the tangent stiffnesses in the model and the data is given by $\frac{\mathrm{d}\mathbf{p}_{\mathrm{mod}}(\boldsymbol{\theta})}{\mathrm{d}r}-\frac{\mathbf{p}_{\mathrm{exp}}(i+1)-\mathbf{p}_{\mathrm{exp}}(i-1)}{r(i+1)-r(i-1)}$ where $\mathbf{p}_{\mathrm{exp}}(i)$ refers to the $i$th data point. The likelihood function proposed is given as 

\begin{equation}
P(\mathbf{p}_{\mathrm{exp}}\mid\mathbf{p}_{\mathrm{mod}}(\boldsymbol{\theta}))=\exp\left(-\frac{(\mathbf{p}_{\mathrm{mod}}(\boldsymbol{\theta})-\mathbf{p}_{\mathrm{exp}})^{2}}{\sigma^{2}}-\frac{\beta}{\sigma^{2}}\left(\frac{\mathrm{d}\mathbf{p}_{\mathrm{mod}}(\boldsymbol{\theta})}{\mathrm{d}r}-\frac{\mathbf{p}_{\mathrm{exp}}(i+1)-\mathbf{p}_{\mathrm{exp}}(i-1)}{r(i+1)-r(i-1)}\right)^{2}\right),\label{eq:LikelihoodFormAssumption}
\end{equation}
where $\frac{1}{\sigma^{2}}$ is the \emph{precision }associated with the errors and $\beta$ is the factor the controls the precision of the error in slope relative to the absolute error. The precision is inversely related to the tolerance in difference between model and data, i.e. lower tolerance for errors $\Rightarrow$ higher precision.

\subsection{Posterior distributions -- Markov Chain Monte Carlo sampling}

Thus our task becomes one of computing $P(\boldsymbol{\theta}\mid\mathbf{p}_{\mathrm{exp}})$ from equation \eqref{eq:BayesRule_modelParams} given that the numerator on the right hand side is equation \eqref{eq:LikelihoodFormAssumption}. 

The denominator of equation \eqref{eq:BayesRule_modelParams} is known as the probability of the evidence, and is given by

\begin{equation}
P(\mathbf{p}_{\mathrm{exp}})=\int\limits _{\boldsymbol{\theta}\in\boldsymbol{\Theta}}P(\mathbf{p}_{\mathrm{exp}}\mid\boldsymbol{\theta})P(\boldsymbol{\theta})\,\mathrm{d}\boldsymbol{\theta}\label{eq:DenominatorIntegral}
\end{equation}

Since this integral is typically very hard to compute, we will use a Monte Carlo algorithm to compute the posterior distribution, without having to explicitly compute the above integral. The algorithm used is known as the Metropolis--Hastings algorithm \cite{metropolis_equation_1953,hastings_monte_1970,gilks_markov_1995}which is used to compute the posterior distribution $P(\boldsymbol{\theta}\mid\mathbf{p}_{\mathrm{exp}})$ and is described in algorithm \eqref{alg:MHalgo}. Further details of this algorithm may be found in \cite{gilks_markov_1995}.

\begin{algorithm}
\begin{algorithmic}[1]
\STATE{Initialize $\boldsymbol{\theta}_{0}$; \textbf{set} $i=0$}

\WHILE{$i<\textsf{MaxIter}$}

\whilebody{\STATE{Sample a point $\boldsymbol{\theta}_{\textrm{cand}}$ according to $q(\cdot\mid\boldsymbol{\theta}_{i})$}\STATE{Find $\mathbf{p}_{\mathrm{mod}}(\boldsymbol{\theta}_{\mathrm{cand}})$with $\boldsymbol{\theta}_{\mathrm{cand}}$ and as the model parameter and $\mathbf{p}_{\mathrm{mod}}(\boldsymbol{\theta}_{i})$ with $\boldsymbol{\theta}_{i}$}

\STATE{Compute likelihood values $P(\mathbf{p}_{\mathrm{exp}}\mid\mathbf{p}_{\mathrm{mod}}(\boldsymbol{\theta}_{\mathrm{cand}}))$ , $P(\mathbf{p}_{\mathrm{exp}}\mid\mathbf{p}_{\mathrm{mod}}(\theta_{i})$ (eqns. \eqref{eq:LikelihoodFormAssumption}) }

\STATE{Compute priors $P(\boldsymbol{\theta}_{\mathrm{cand}})$ , $P(\boldsymbol{\theta}_{i})$ (The prior is uniform; these probabilities are equal) }

\STATE{Compute value of acceptance criterion $\alpha(\boldsymbol{\theta}_{i}\mid\boldsymbol{\theta}_{\textrm{cand}})$ }

\STATE{Sample a uniform random variable $u\sim\mathrm{U}(0,1)$}

\IF{$u\leq\alpha(\boldsymbol{\theta}_{i}\mid\boldsymbol{\theta}_{\textrm{cand}})$}

\ifbody{\STATE{$\boldsymbol{\theta}_{i+1}=\boldsymbol{\theta}_{\textrm{cand}}$}

\ELSE{}

\STATE{$\boldsymbol{\theta}_{i+1}=\boldsymbol{\theta}_{i}$}}\STATE{\textbf{set} $i=i+1$}}
\end{algorithmic}

\caption{Metropolis--Hastings algorithm}
\label{alg:MHalgo}
\end{algorithm}

\section{Results \label{sec:Results-and-Discussion}}

The probability distribution of the model parameters are presented for a subset of the data that is shown in figure \ref{fig:Inflation-experimental-data}. This choice is in order to demonstrate an application of the probability distributions to a classification problem. Towards this, the samples are grouped into two classes - samples 1,2 and 3 into class $C_{1}$ and samples 4 and 5 into class $C_{2}$. For each of the classes, the data $\mathbf{p}_{\mathrm{exp}}$ is chosen as two of the three experiments per sample. Therefore, for class $C_{1}$, $\mathbf{p}_{\mathrm{exp}}$ is the set of six experiments (shown in figure \ref{fig:ExpData_Classes} (a)) and for class $C_{2}$, $\mathbf{p}_{\mathrm{exp}}$ is the set of four experiments (shown in figure \ref{fig:ExpData_Classes}(b)). The probability distribution for the parameters for the two classes corresponding to each of the strain energy functions from section \eqref{sub:Forms-of-constitutive} are shown in figure \ref{fig:ModelParams_ProbDist}. 

A cursory inspection of the probability distributions show that the regions of the parameter space with high probability are not the same for the two classes. Qualitatively, this observation shows that the two classes that we considered are `different'. Indeed, a quantitative measure of this difference, as mentioned earlier, could be obtained by computing a distance such as the K-L divergence, between the probability distributions.

Another noteworthy feature in the model parameter probability distributions in figure \ref{fig:ModelParams_ProbDist} is that, for the Ogden model and the Criscione-type model, the probability distributions are bimodal. For the Ogden model, this is due to the fact that in equation \ref{eq:OgdenModel}, the strain energy is symmetric with respect to the sign of the exponent parameter $\alpha$. It can be seen that this symmetry is reflected in the two modes in each of the contour plots in figure \ref{fig:ModelParams_ProbDist} (c) and (d). Similarly, for the Criscione-type model, the modes are due to the hyperbolic cosine function in the strain energy, since $\cosh$ is an even function. This is reflected in the symmetry in parameter $b$ in figure \ref{fig:ModelParams_ProbDist} (e) and (f).$ $

Thus, the experimental data is susceptible to two groups of explanations, given the particular choice of Ogden and Criscione-type models.

\begin{center}
\begin{figure}[H]
\begin{centering}
\subfloat[Class $C_{1}$]{\centering{}\includegraphics[width=0.55\textwidth]{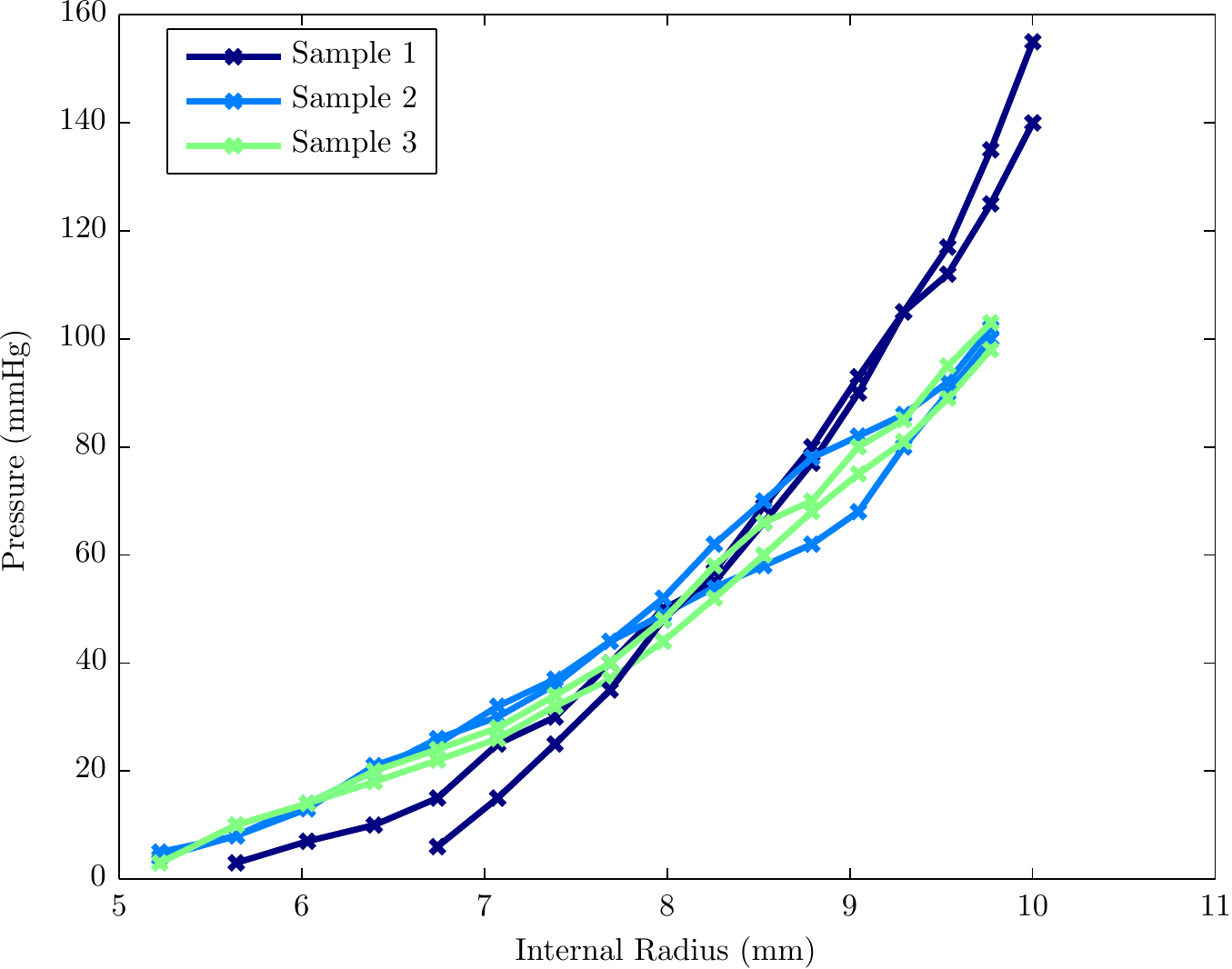}} \subfloat[Class $C_{2}$]{\centering{}\includegraphics[width=0.55\textwidth]{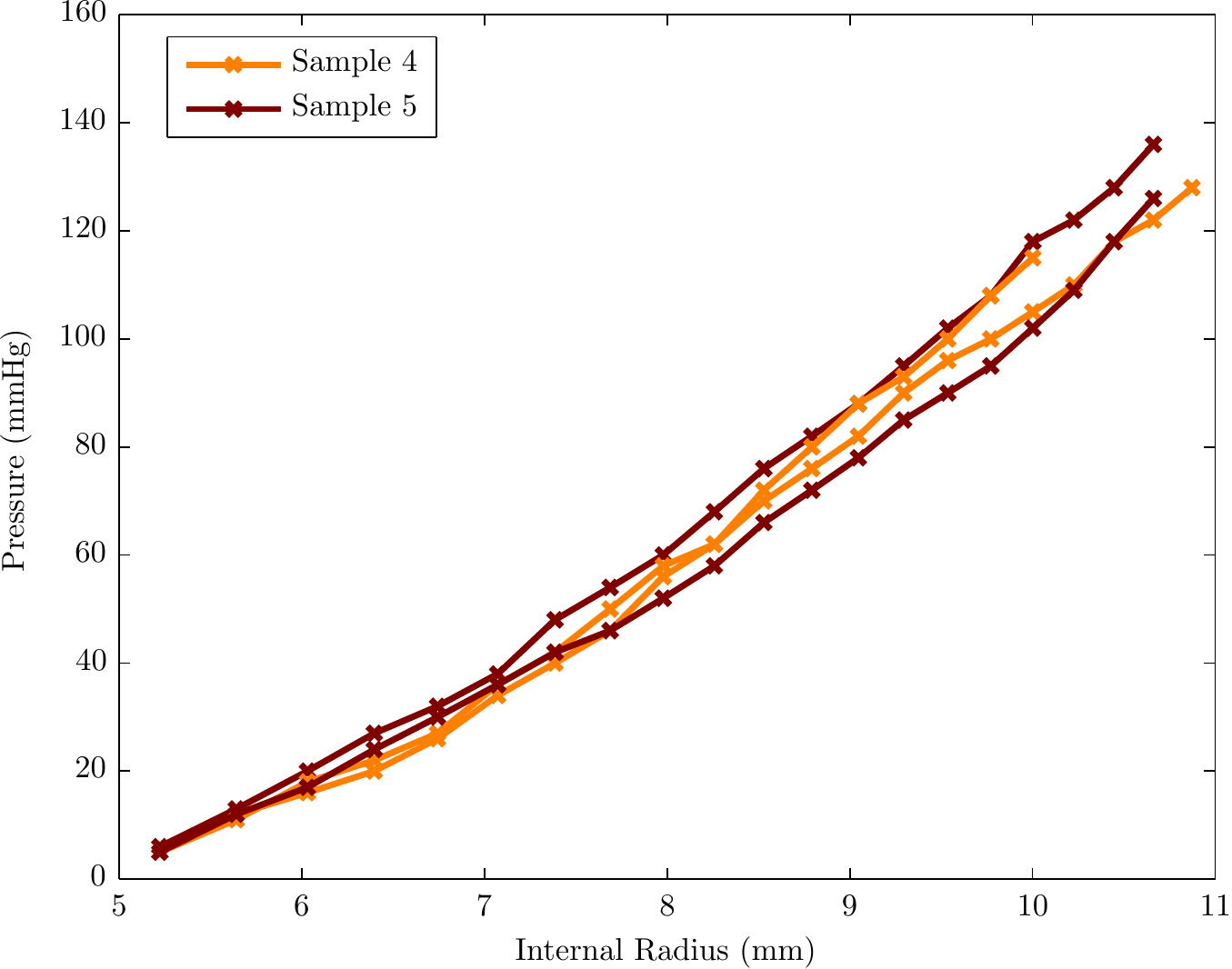}}
\par\end{centering}

\caption{The experimental data shown as (a) class $C_{1}$ and class (b) class $C_{2}$. These two datasets are used to compute the probability distributions shown in figure \ref{fig:ModelParams_ProbDist}.\label{fig:ExpData_Classes}}
\end{figure}

\par\end{center}

\begin{center}
\begin{figure}[H]
\begin{centering}
\subfloat[Samples 1,2,3 -- Power-law neo-Hookean]{\begin{centering}
\includegraphics[width=0.55\textwidth]{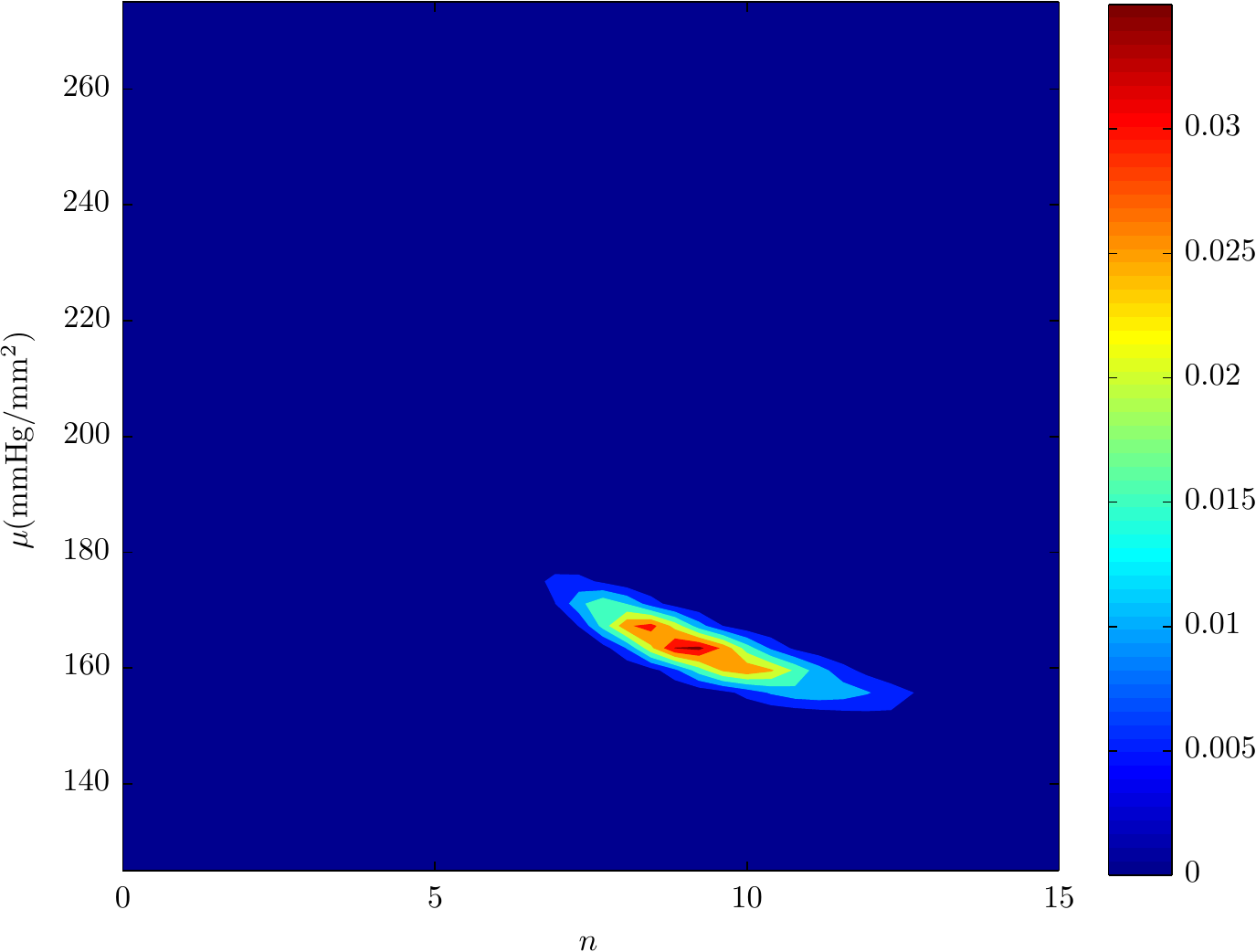}
\par\end{centering}

}\enspace{}\subfloat[Samples 4,5 -- Power-law neo-Hookean]{\centering{}\includegraphics[width=0.55\textwidth]{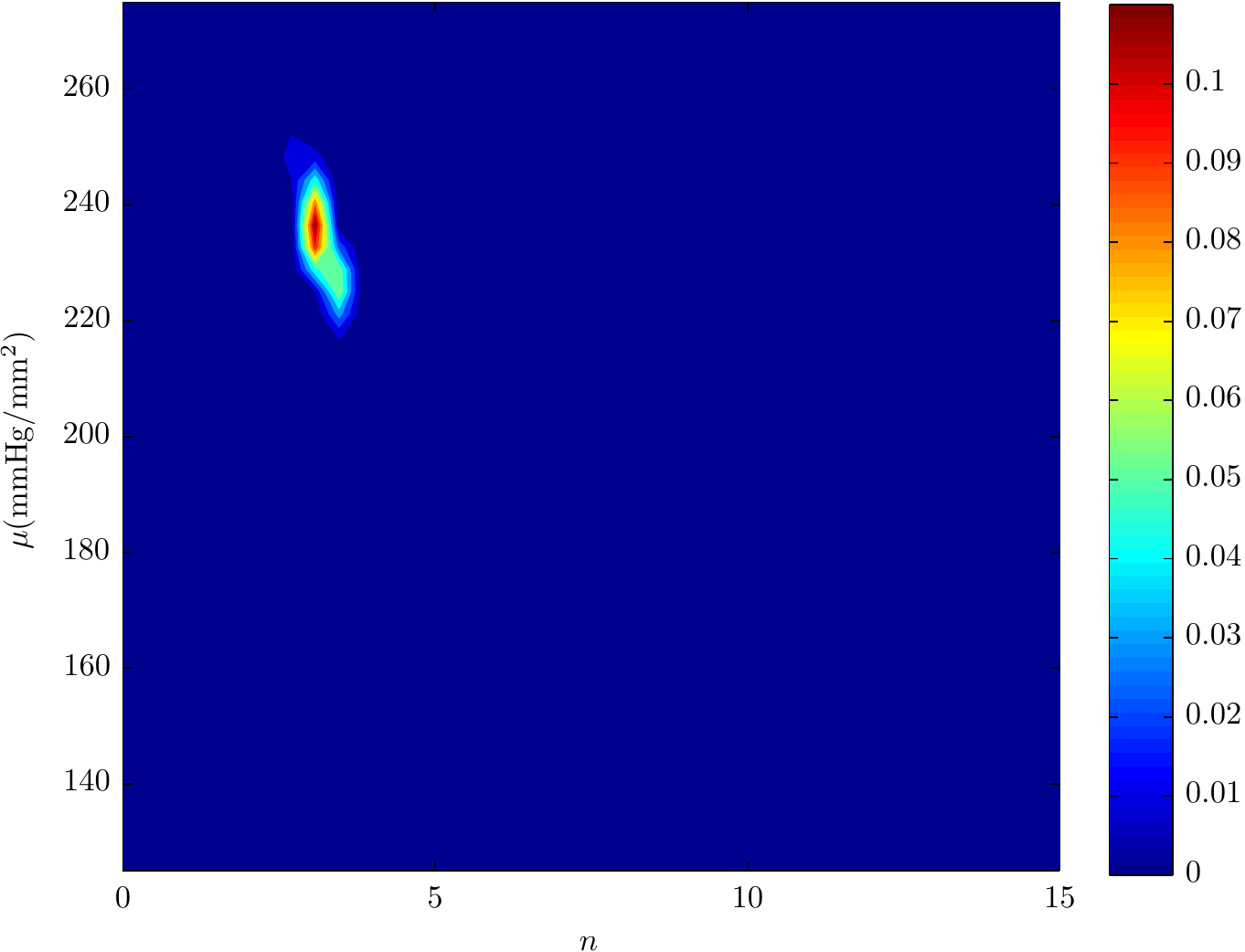}}
\par\end{centering}

\begin{centering}
\subfloat[Samples 1,2,3 -- Ogden strain energy]{\centering{}\includegraphics[width=0.55\textwidth]{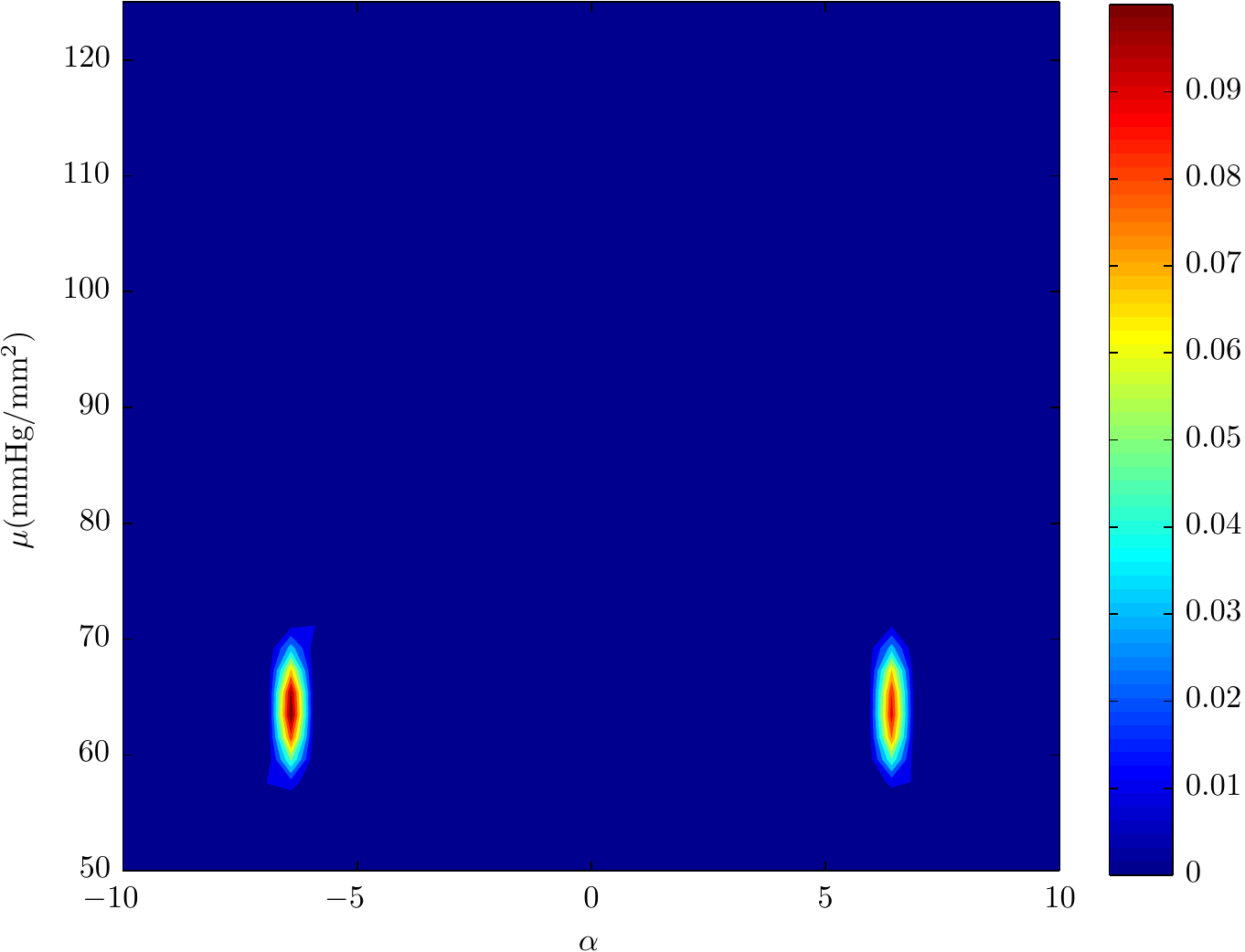}}\enspace{}\subfloat[Samples 4,5 -- Ogden strain energy]{\centering{}\includegraphics[width=0.55\textwidth]{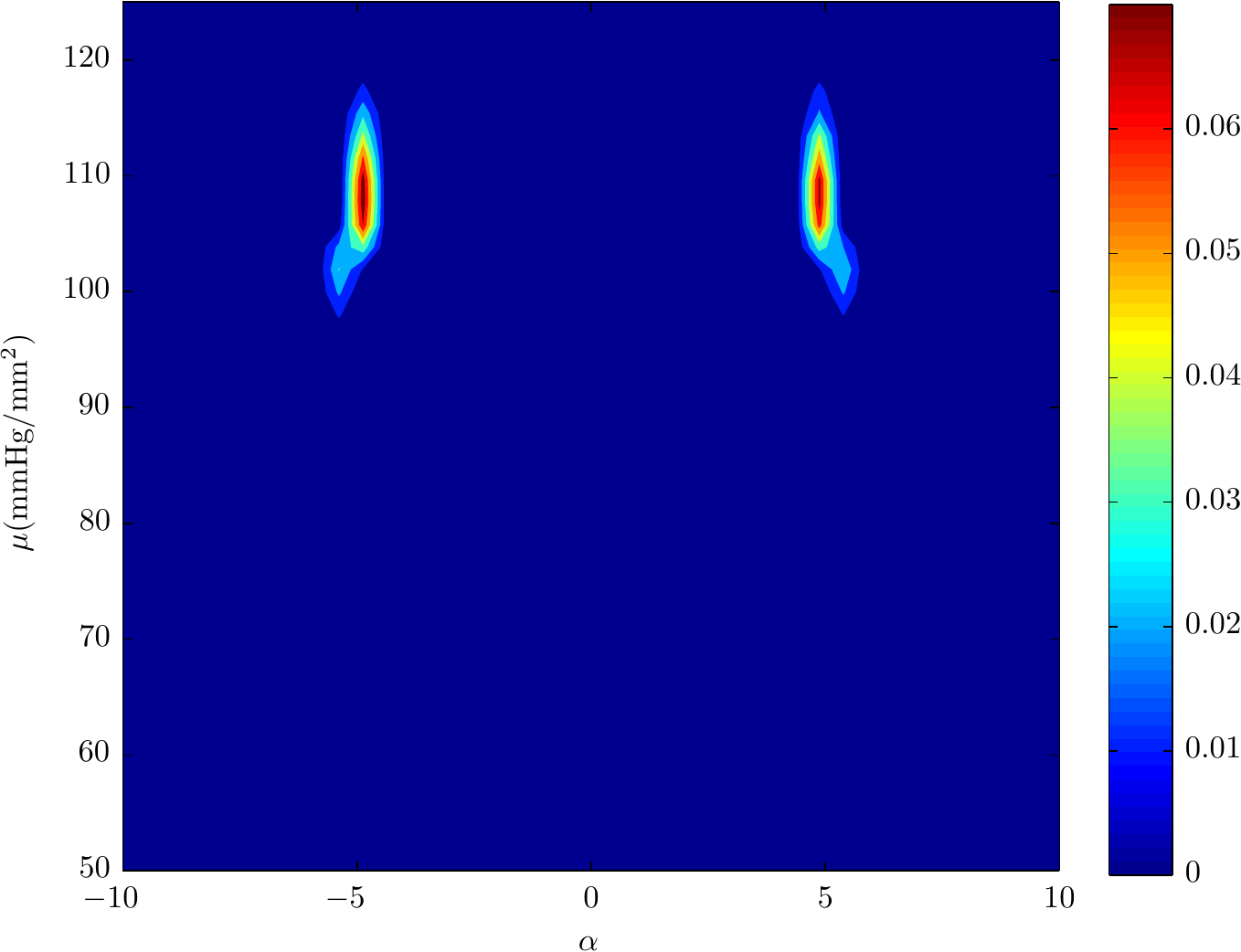}}
\par\end{centering}

\begin{centering}
\subfloat[Samples 1,2,3 -- Criscione-type strain energy]{\centering{}\includegraphics[width=0.55\textwidth]{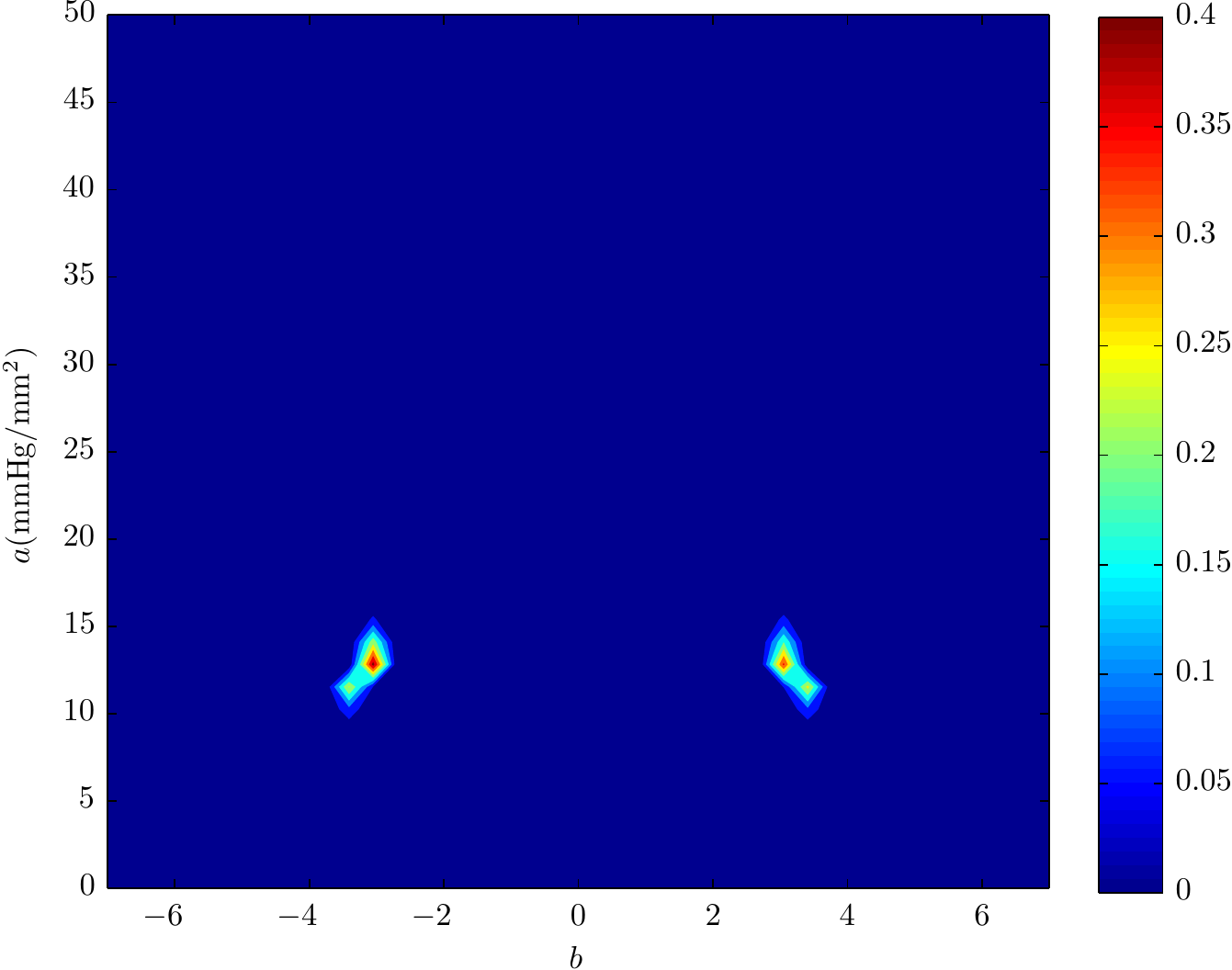}}\enspace{}\subfloat[Samples 4,5 -- Criscione-type strain energy]{\centering{}\includegraphics[width=0.55\textwidth]{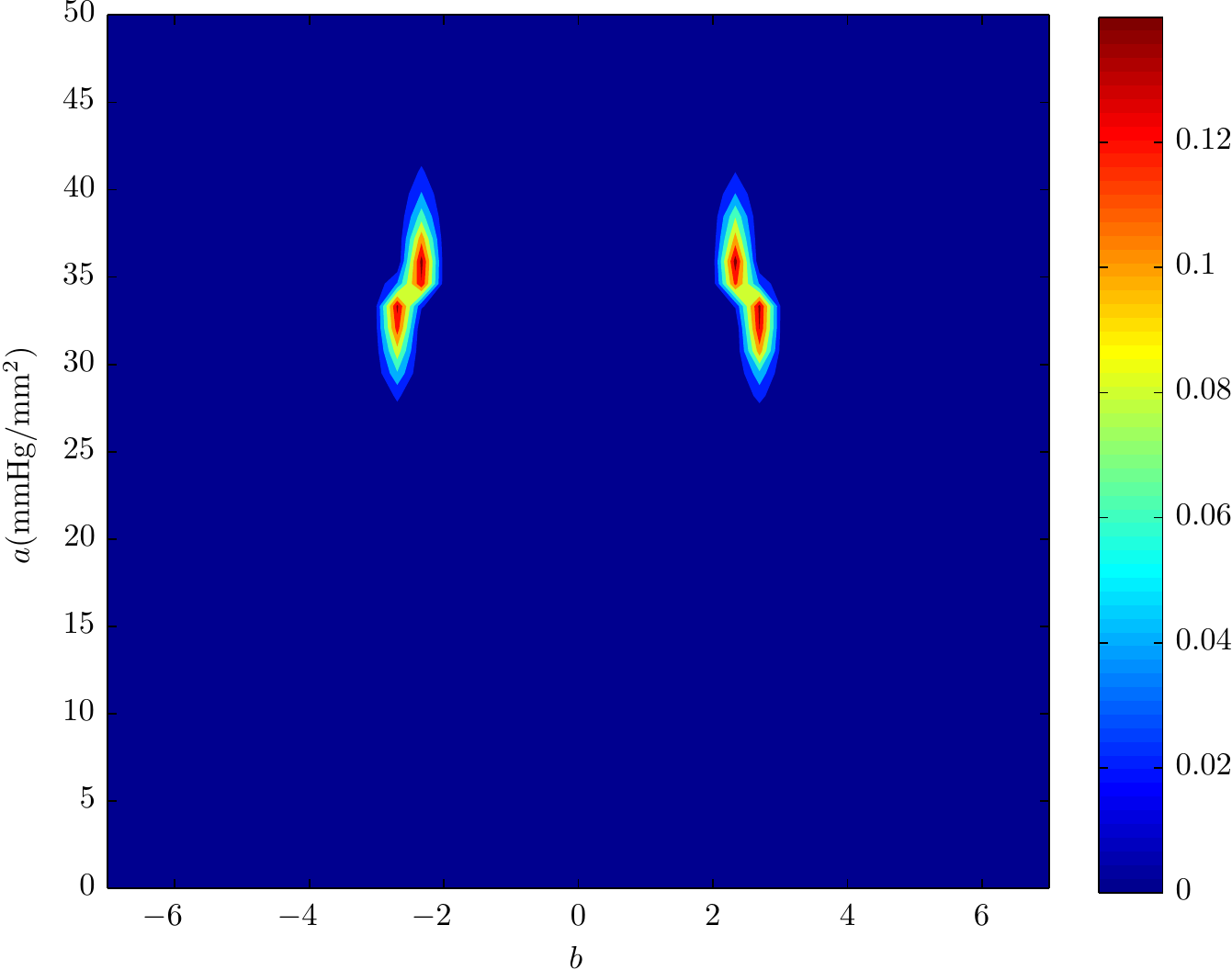}}
\par\end{centering}

\caption{The probability distributions of the model parameters of the three strain energies using two out of three experiments for each class (see section \ref{sec:Results-and-Discussion}) as data for each are shown.\label{fig:ModelParams_ProbDist}}

\end{figure}

\par\end{center}

\section{Using model parameter distributions for classification\label{sec:Using-model-parameter}}

An interesting application of the probability distributions for the model parameters is for classification problems. The techniques used for classification are typically of two types: deterministic or probabilistic. Deterministic algorithms describe hard boundaries between groups of observations, thus dividing the entire set of observations into clearly structured groups. Examples of such techniques are hypothesis testing, $k$ -means clustering and computing separating hyperplanes using support vector machines. On the other hand, probabilistic algorithms relax such hard boundaries and instead assign the probability that a particular observation belonging to a certain class. Probabilistic techniques thus utilize a ``fuzzy'' classification approach. Examples of probabilistic algorithms for classification include logistic regression \cite{hosmer_applied_2013} and neural networks \cite{zhang_neural_2000}. Probabilistic classification allows for naturally incorporating any uncertainty thus allowing for making decision only if sufficiently confident. This feature is especially critical in medical applications and thus probabilistic techniques have been the favored approach for applications in medical diagnosis \cite{croft_mathematical_1974,kononenko_machine_2001,ledley_reasoning_1959,szolovits_categorical_1978,warner_hr_mathematical_1961}.

The Bayesian framework naturally allows for assigning probabilities to each of the classes considered using the probability distributions computed earlier in this paper. Classifying the data \emph{not} used as part of $\mathbf{p}_{\mathrm{exp}}$ into classes $C_{1}$ and $C_{2}$ (see figure \ref{fig:ExpData_Classes}) is demonstrated as an example of the approach.

\subsection{Class membership probabilities\label{sub:Class-membership-probabilities}}

Let classes $C_{1},\ldots,C_{n}$ be classes into which a newly obtained data set $\mathbf{p}_{\mathrm{new}}$ is to be classified. The classes $C_{i}$ are each associated with a probability distribution for the model parameter $\boldsymbol{\theta}$, $P(\boldsymbol{\theta}\mid C_{i})$. The probability distributions $P(\boldsymbol{\theta}\mid C_{i})$ are obtained through a Bayesian inference procedure on the training data $\mathbf{p}_{\mathrm{exp}}$ (see section \ref{sub:Probability-distribution-of}).

For the classification problem, given a newly obtained data set $\mathbf{p}_{\mathrm{new}}$, the probability that it belongs to a class $C_{i}$, $P(C_{i}\mid\mathbf{p}_{\mathrm{new}})$ can be computed using the Bayes' rule as,

\begin{align}
P(C_{i}\mid\mathbf{p}_{\mathrm{new}}) & =\frac{P(\mathbf{p}_{\mathrm{new}}\mid C_{i})P(C_{i})}{P(\mathbf{p}_{\mathrm{new}})}\label{eq:BayesRule_classification}
\end{align}

In equation \ref{eq:BayesRule_classification}, the probability on the left hand side $P(C_{i}\mid\mathbf{p}_{\mathrm{new}})$ is the \emph{posterior }probability, $P(\mathbf{p}_{\mathrm{new}}\mid C_{i})$ is called the \emph{likelihood} function and $P(C_{i})$ is the \emph{prior }probability.\emph{ }

The likelihood, $P(\mathbf{p}_{\mathrm{new}}\mid i)$ is also known as the \emph{marginal likelihood }due to the marginalization of $P(\mathbf{p}_{\mathrm{new}}\mid i,\boldsymbol{\theta})$ over the parameter space $\boldsymbol{\Theta}$ (i.e.) 

\begin{align*}
P(\mathbf{p}_{\mathrm{new}}\mid C_{i}) & =\int_{\boldsymbol{\Theta}}P(\mathbf{p}_{\mathrm{new}}|\boldsymbol{\theta},C_{i})P(\boldsymbol{\theta}\mid C_{i})\,\mathrm{d}\boldsymbol{\theta}
\end{align*}

where, $ $$P(\mathbf{p}_{\mathrm{new}}|\boldsymbol{\theta},C_{i})$ is the likelihood associated with observing the data $\mathbf{p_{\mathrm{new}}}$ given that the model parameters are \textbf{$\boldsymbol{\theta}$} (which is the same as in equation \eqref{eq:LikelihoodFormAssumption})\textbf{ }and $P(\boldsymbol{\theta}\mid C_{i})$ is the probability distribution on the parameter $\boldsymbol{\theta}$ associated with the class $C_{i}$.

Equation \ref{eq:BayesRule_classification}

\begin{equation}
P(C_{i}\mid\mathbf{p}_{\mathrm{new}})=\frac{\int_{\boldsymbol{\Theta}}P(\mathbf{p}_{\mathrm{new}}|\boldsymbol{\theta},i)P(\boldsymbol{\theta}\mid i)\,\mathrm{d}\boldsymbol{\theta}\, P(i)}{P(\mathbf{p}_{\mathrm{new}})}\label{eq:BayesRule_Classification_Marginalization}
\end{equation}

Figure \ref{fig:ClassProb_Hist} shows the results of this classification applied to the data set shown in figure \ref{fig:NewData_Classes}. 

\begin{center}
\begin{figure}[H]
\begin{centering}
\includegraphics[scale=0.75]{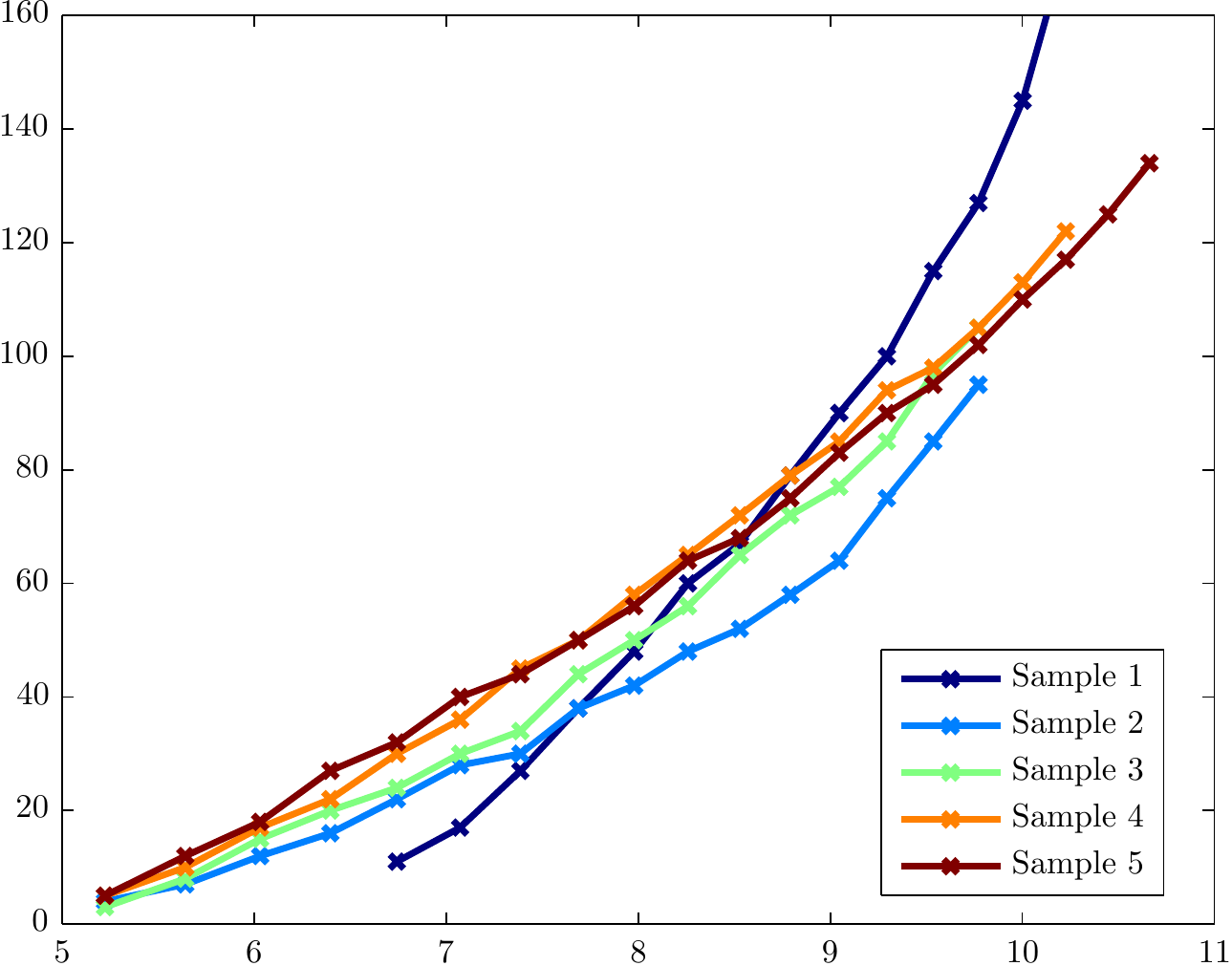}
\par\end{centering}

\caption{The data, $\mathbf{p}_{\mathrm{new}}$, used for classification (see section \ref{sub:Class-membership-probabilities}). Each of the classes $C_{1}$ and $C_{2}$ experiments plotted here are assigned a probability (see figure \ref{fig:ClassProb_Hist}). \label{fig:NewData_Classes}}
\end{figure}

\par\end{center}

\begin{center}
\begin{figure}[H]
\begin{centering}
\subfloat[Power law neo-Hookean model]{\includegraphics[width=0.6\textwidth]{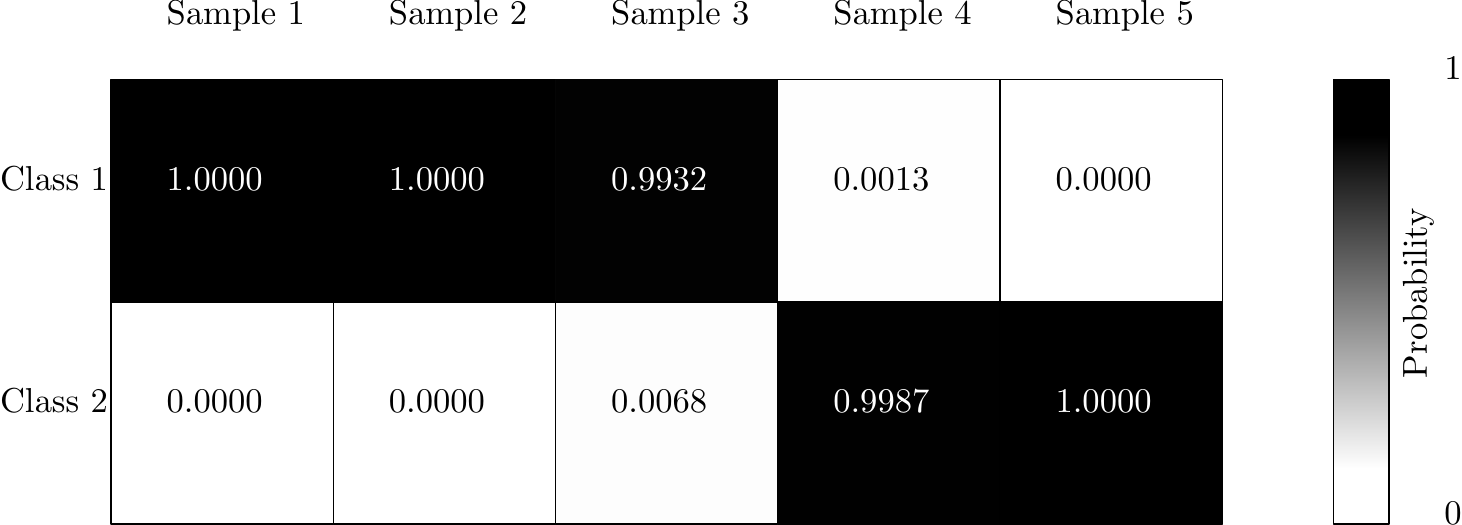}}
\par\end{centering}

\begin{centering}
\subfloat[Ogden model]{\includegraphics[width=0.6\textwidth]{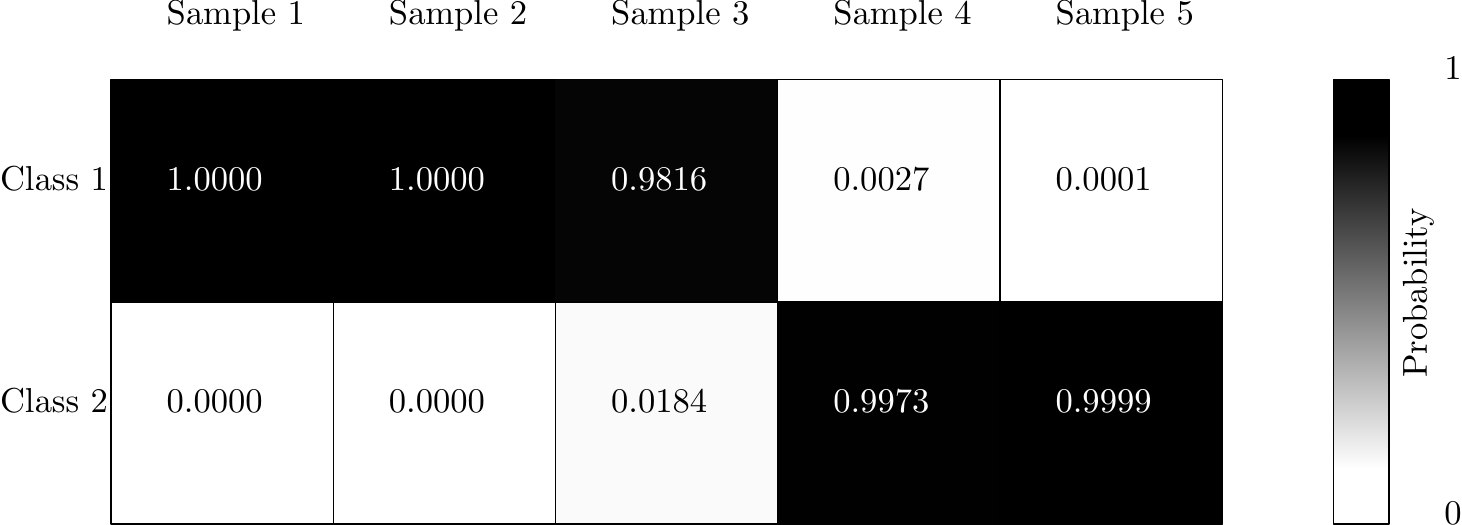}}
\par\end{centering}

\begin{centering}
\subfloat[Criscione-type model]{\includegraphics[width=0.6\textwidth]{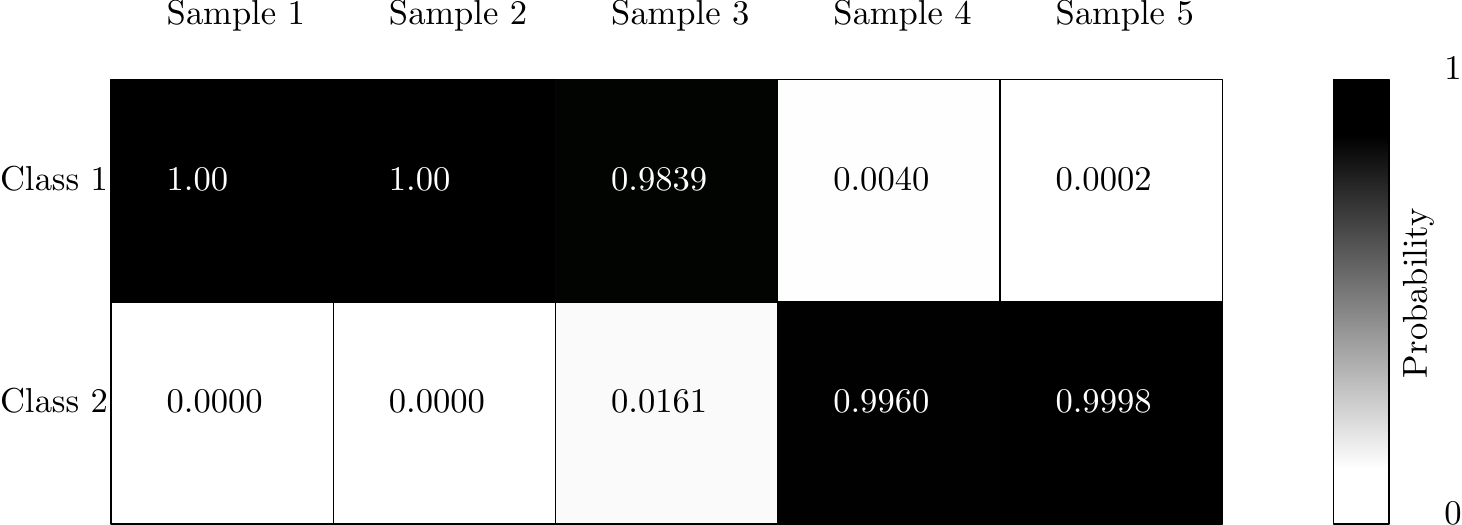}}
\par\end{centering}

\caption{The class probability histograms for the data set $\mathbf{p}_{\mathrm{new}}$ and classified between class $C_{1}$ (class of samples 1,2,3) and $C_{2}$(class of samples 4,5). The row headings indicate the source of the data $\mathbf{p}_{\mathrm{new}}$, although this information is not used in the classification procedure (section \ref{sub:Class-membership-probabilities}). Each of the figures correspond to the model (strain energy functions) used in the marginal likelihood. \label{fig:ClassProb_Hist}}

\end{figure}

\par\end{center}

\section{Summary }

We have developed a method for representing the response of biomaterials in which the model parameters are themselves considered as probability distributions. This approach treats the distribution of responses of nominally similar biomaterials as a feature and not as something that can be eliminated by careful control of experiments. By a systematic use of Bayesian inference, the approach can be extended to situations in which model parameter distributions can be improved as and when more data is available. In other words, the approach allows the modeler to ``learn'' more about the parameters as additional data becomes available. Furthermore, we have also presented a new method for model based classification of data based on parameters that were obtained from training data.

\bibliographystyle{elsarticle-num}
\bibliography{references}

\begin{thebibliography}{10}
\expandafter\ifx\csname url\endcsname\relax
  \def\url#1{\texttt{#1}}\fi
\expandafter\ifx\csname urlprefix\endcsname\relax\def\urlprefix{URL }\fi
\expandafter\ifx\csname href\endcsname\relax
  \def\href#1#2{#2} \def\path#1{#1}\fi

\bibitem{mollica_modeling_2007}
F.~Mollica, L.~Preziosi, K.~R. Rajagopal, Modeling of Biological Materials
  (Modeling and Simulation in Science, Engineering and Technology),
  Birkh\"auser, 2007.

\bibitem{humphrey_cardiovascular_2002}
J.~D. Humphrey, Cardiovascular Solid Mechanics: Cells, Tissues, and Organs,
  2002nd Edition, Springer, 2002.

\bibitem{fung_biomechanics:_1993}
Y.~C. Fung, Biomechanics: Mechanical Properties of Living Tissues, Second
  Edition, 2nd Edition, Springer, 1993.

\bibitem{holzapfel_mechanics_2006}
G.~A. Holzapfel, R.~W. Ogden, Mechanics of Biological Tissue, 2006th Edition,
  Springer, 2006.

\bibitem{van_andel_mechanical_2003}
C.~J. van Andel, P.~V. Pistecky, C.~Borst, Mechanical properties of porcine and
  human arteries: implications for coronary anastomotic connectors, The Annals
  of Thoracic Surgery 76~(1) (2003) 58--64.

\bibitem{vande_geest_effects_2006}
J.~P. Vande~Geest, M.~S. Sacks, D.~A. Vorp, The effects of aneurysm on the
  biaxial mechanical behavior of human abdominal aorta, Journal of Biomechanics
  39~(7) (2006) 1324--1334.

\bibitem{garcia-herrera_mechanical_2012}
C.~M. Garc\'ia-Herrera, D.~J. Celentano, M.~A. Cruchaga, F.~J. Rojo, J.~M.
  Atienza, G.~V. Guinea, J.~M. Goicolea, Mechanical characterisation of the
  human thoracic descending aorta: experiments and modelling, Computer Methods
  in Biomechanics and Biomedical Engineering 15~(2) (2012) 185--193.

\bibitem{carboni_passive_2007}
M.~Carboni, G.~W. Desch, H.~W. Weizs\"acker, Passive mechanical properties of
  porcine left circumflex artery and its mathematical description, Medical
  Engineering \& Physics 29~(1) (2007) 8--16.

\bibitem{holzapfel_determination_2005}
G.~A. Holzapfel, G.~Sommer, C.~T. Gasser, P.~Regitnig, Determination of
  layer-specific mechanical properties of human coronary arteries with
  nonatherosclerotic intimal thickening and related constitutive modeling,
  American Journal of Physiology - Heart and Circulatory Physiology 289~(5)
  (2005) H2048--H2058.

\bibitem{paul_andersohn_modeling_2013}
A.~Paul~Andersohn, Modeling frameworks for representing the mechanical behavior
  of tissues with a specific look at vasculature, Master's thesis, Texas {A\&M}
  University, College Station (Dec. 2013).

\bibitem{ogden_fitting_2004}
R.~W. Ogden, G.~Saccomandi, I.~Sgura, Fitting hyperelastic models to
  experimental data, Computational Mechanics 34~(6) (2004) 484--502.

\bibitem{ogden_non_1997}
R.~W. Ogden, Non-linear elastic deformations, Dover Publications, 1997.

\bibitem{raghavan_toward_2000}
M.~Raghavan, D.~A. Vorp, Toward a biomechanical tool to evaluate rupture
  potential of abdominal aortic aneurysm: identification of a finite strain
  constitutive model and evaluation of its applicability, Journal of
  Biomechanics 33~(4) (2000) 475--482.

\bibitem{criscione_constitutive_2004}
J.~C. Criscione, A constitutive framework for tubular structures that enables a
  semi-inverse solution to extension and inflation, Journal of Elasticity
  77~(1) (2004) 57--81.

\bibitem{srinivasa_use_2012}
A.~Srinivasa, On the use of the upper triangular (or qr) decomposition for
  developing constitutive equations for green-elastic materials, International
  Journal of Engineering Science 60 (2012) 1--12.

\bibitem{gosling_terminology_2003}
R.~G. Gosling, M.~M. Budge, Terminology for describing the elastic behavior of
  arteries, Hypertension 41~(6) (2003) 1180--1182, {PMID:} 12756217.

\bibitem{metropolis_equation_1953}
N.~Metropolis, A.~W. Rosenbluth, M.~N. Rosenbluth, A.~H. Teller, E.~Teller,
  Equation of state calculations by fast computing machines, The Journal of
  Chemical Physics 21~(6) (1953) 1087--1092.

\bibitem{hastings_monte_1970}
W.~K. Hastings, Monte carlo sampling methods using markov chains and their
  applications, Biometrika 57~(1) (1970) 97--109.

\bibitem{gilks_markov_1995}
W.~R. Gilks, S.~Richardson, D.~Spiegelhalter, Markov Chain Monte Carlo in
  Practice (Chapman \& {Hall/CRC} Interdisciplinary Statistics), 1st Edition,
  Chapman and {Hall/CRC}, 1995.

\bibitem{hosmer_applied_2013}
D.~W. Hosmer, S.~Lemeshow, R.~X. Sturdivant, Applied Logistic Regression (Wiley
  Series in Probability and Statistics), 3rd Edition, Wiley, 2013.

\bibitem{zhang_neural_2000}
G.~Zhang, Neural networks for classification: a survey, {IEEE} Transactions on
  Systems, Man, and Cybernetics, Part C: Applications and Reviews 30~(4) (2000)
  451--462.

\bibitem{croft_mathematical_1974}
D.~J. Croft, R.~E. Machol, Mathematical methods in medical diagnosis, Annals of
  Biomedical Engineering 2~(1) (1974) 69--89.

\bibitem{kononenko_machine_2001}
I.~Kononenko, Machine learning for medical diagnosis: history, state of the art
  and perspective, Artificial Intelligence in Medicine 23~(1) (2001) 89--109.

\bibitem{ledley_reasoning_1959}
R.~S. Ledley, L.~B. Lusted, Reasoning foundations of medical diagnosis symbolic
  logic, probability, and value theory aid our understanding of how physicians
  reason, Science 130~(3366) (1959) 9--21, {PMID:} 13668531.

\bibitem{szolovits_categorical_1978}
P.~Szolovits, S.~G. Pauker, Categorical and probabilistic reasoning in medical
  diagnosis, Artificial Intelligence 11~(1-2) (1978) 115--144.

\bibitem{warner_hr_mathematical_1961}
{Warner {HR}}, {Toronto {AF}}, {Veasey L}, {Stephenson R}, A mathematical
  approach to medical diagnosis: Application to congenital heart disease, The
  Journal of American Medical Association 177~(3) (1961) 177--183.

\end{thebibliography}

\end{document}